\documentclass[journal]{IEEEtran}
\IEEEoverridecommandlockouts
\usepackage{amsfonts}
\usepackage{amssymb}
\usepackage[cmex10]{amsmath}
\usepackage{graphicx}
\usepackage{subfigure}
\usepackage{verbatim}
\usepackage{cite}
\usepackage[labelfont=bf]{caption}
\usepackage{diagbox}
\usepackage{tabularx,booktabs}
\usepackage{caption}
\usepackage{amsthm}
\usepackage{booktabs}
\usepackage[ruled,lined]{algorithm2e}
\usepackage{multirow}
\usepackage{setspace}
\usepackage{stfloats}
\usepackage{fancyhdr}
\usepackage{float}
\usepackage{algorithmic}
\usepackage{enumerate}
\usepackage{amsmath}
\usepackage[ruled]{algorithm2e}
\usepackage{subfigure}
\usepackage{array}
\usepackage[usenames,dvipsnames]{color}

\usepackage{calc}
\usepackage{booktabs}
\usepackage{tabularx}
\usepackage[acronym,nomain]{glossaries}

\makeglossaries

\newacronym{led}{LED}{Light-Emitting Diode}
\newacronym{vcc}{VCC}{positive Versorgungsspannung}
\newacronym{gnd}{GND}{Ground, negative Versorgungsspannung}

\newlength\maxlength
\newlength\thislength

\newglossarystyle{mystyle}
{%
  {
    \setlength{\maxlength}{0pt}%
    \forglsentries[\currentglossary]{\thislabel}%
    {%
       \settowidth{\thislength}{\glsentryshort{\thislabel}}%
       \ifdim\thislength>\maxlength
         \setlength{\maxlength}{\thislength}%
       \fi
    }%
    \settowidth{\thislength}{\hspace{1.5em}=\hspace{1em}}%
    \setlength{\glsdescwidth}{\linewidth-\maxlength-\thislength-2\tabcolsep}%
    \begin{tabular}{l@{\hspace{1.5em}=\hspace{1em}}p{\glsdescwidth}}
    \toprule
    \multicolumn{1}{l}{\textbf{Abk.}} &
    \multicolumn{1}{@{}l}{\textbf{Bedeutung}}\\%
    \midrule
  }%
  {
     \bottomrule
     \end{tabular}%
  }%
  \renewcommand*{\glsgroupheading}[1]{}%
}

\def\BibTeX{{\rm B\kern-.05em{\sc i\kern-.025em b}\kern-.08em
    T\kern-.1667em\lower.7ex\hbox{E}\kern-.125emX}}
\begin{document}
\thispagestyle{empty}
\pagestyle{empty}
\newcommand{\be}{\begin{equation}}
\newcommand{\ee}{\end{equation}}
\newcommand{\br}{{\mbox{\boldmath{$r$}}}}
\newcommand{\bp}{{\mbox{\boldmath{$p$}}}}
\newcommand{\bpi}{\mbox{\boldmath{ $\pi $}}}
\newcommand{\bn}{{\mbox{\boldmath{$n$}}}}
\newcommand{\balfa}{{\mbox{\boldmath{$\alpha$}}}}
\newcommand{\ba}{\mbox{\boldmath{$a $}}}
\newcommand{\bta}{\mbox{\boldmath{$\beta $}}}
\newcommand{\bg}{\mbox{\boldmath{$g $}}}
\newcommand{\bPsi}{\mbox{\boldmath{$\Psi $}}}
\newcommand{\bsigma}{\mbox{\boldmath{ $\Sigma $}}}
\newcommand{\bGamma}{{\bf \Gamma }}
\newcommand{\bA}{{\bf A }}
\newcommand{\bP}{{\bf P }}
\newcommand{\bX}{{\bf X }}
\newcommand{\bI}{{\bf I }}
\newcommand{\bR}{{\bf R }}
\newcommand{\bZ}{{\bf Z }}
\newcommand{\bz}{{\bf z }}
\newcommand{\bx}{{\mathbf{x}}}
\newcommand{\bM}{{\bf M}}
\newcommand{\bU}{{\bf U}}
\newcommand{\bD}{{\bf D}}
\newcommand{\bJ}{{\bf J}}
\newcommand{\bH}{{\bf H}}
\newcommand{\bK}{{\bf K}}
\newcommand{\bm}{{\bf m}}
\newcommand{\bN}{{\bf N}}
\newcommand{\bC}{{\bf C}}
\newcommand{\bL}{{\bf L}}
\newcommand{\bF}{{\bf F}}
\newcommand{\bv}{{\bf v}}
\newcommand{\bSigma}{{\bf \Sigma}}
\newcommand{\bS}{{\bf S}}
\newcommand{\bs}{{\bf s}}
\newcommand{\bO}{{\bf O}}
\newcommand{\bQ}{{\bf Q}}
\newcommand{\btr}{{\mbox{\boldmath{$tr$}}}}
\newcommand{\bNSCM}{{\bf NSCM}}
\newcommand{\barg}{{\bf arg}}
\newcommand{\bmax}{{\bf max}}
\newcommand{\test}{\mbox{$
\begin{array}{c}
\stackrel{ \stackrel{\textstyle H_1}{\textstyle >} } { \stackrel{\textstyle <}{\textstyle H_0} }
\end{array}
$}}
\newcommand{\tabincell}[2]{\begin{tabular}{@{}#1@{}}#2\end{tabular}}
\newtheorem{Def}{Definition}
\newtheorem{Pro}{Proposition}
\newtheorem{Exa}{Example}
\newtheorem{Rem}{Remark}
\newtheorem{Cor}{Corollary}
\newtheorem{Lem}{Lemma}
\renewcommand{\labelitemi}{$\bullet$}

\title{ Multiple Model Poisson Multi-Bernoulli Mixture Filter for Maneuvering Targets}
\author{\IEEEauthorblockN{Guchong Li}
\thanks{

G. Li is with the School of Information and Communication Engineering, University of Electronic Science and Technology of China, Chengdu 611731, China (e-mail: guchong.li@hotmail.com).
}
}
\maketitle
\begin{abstract}
The Poisson multi-Bernoulli mixture (PMBM) filter is conjugate prior composed of the union of a Poisson point process (PPP) and a multi-Bernoulli mixture (MBM). In this paper, a new PMBM filter for tracking multiple targets with randomly time-varying dynamics under multiple model (MM) is considered. The proposed MM-PMBM filter uses extends the single-model PMBM filter recursion to multiple motion models by using the jump Markov system (JMS). The performance of the proposed algorithm is examined and compared with the MM-MB filter. The simulation results demonstrate that the proposed MM-PMBM filter outperforms the MM-MB filter in terms of the tracking accuracy, including the target states and cardinality, especially for the scenerio with low detection probability. Moreover, the comparisons for the variations of detection probability and standard derivation of measurement noise are also tested via simulation experiments.
\end{abstract}
\IEEEpeerreviewmaketitle

\section{Introduction}
With the increasingly complex monitoring environments, multi-target tracking (MTT) has caused widespread concerns, and has been adopted in many fields \cite{Blackman2004,Garcia2009,Petrovskaya2009,Giorgio2017,Guchong2019}. The goal of MTT is to jointly estimate the target states and time-varying target cardinality from the noisy measurements. For this problem, it is well-known that there are always two major challenges, measurement-origin uncertainty and target motion uncertainty.

In terms of the measurement-origin uncertainty, the core is to solve the association problem between the targets and measurements. The mostly common known detection and tracking algorithms include joint probabilistic data association (JPDA) ~\cite{BarShalom1988,Formann1983} and multiple hypothesis tracking (MHT) ~\cite{Reid1980,Streit1994,MHT} as well as the relatively new finite set statistics (FISST) theory \cite{MahlerPHD,Mahlerbook}. Recently, there is a great concern for the FISST theory, and a series of filters are proposed in order to overcome the unattainable implementation of multi-target Bayes filter, such as probability hypothesis density (PHD) \cite{MahlerPHD}, Cardinality-PHD (CPHD) \cite{MahlerCPHD}, multi-Bernoulli (MB) \cite{VoMB1,Reza2013}, labeled MB (LMB and GLMB) \cite{VoGLMB,VoGLMB2,Reuter,Fantacci2016}, and the recently Poisson MB mixture (PMBM) \cite{Williams2015,Angel2018} filters.

Comparing to the other unlabeled RFS-based filters (PHD, CPHD, and MB), an unique and important characteristic of the PMBM filter is the conjugacy property like GLMB filter \cite{VoGLMB}, which means that the posterior distribution has the same functional form as the prior. Moreover, it is also verified that the PMBM filter has the attractive performance in some scenarios with low detection probability \cite{Williams2015,Xia2017,Angel2018}. Furthermore, two related implementation methods are also proposed, sequential monte carlo (SMC) \cite{Kropfreiter} and Gaussian mixture (GM) \cite{Angel2018}. As a result, the PMBM filter has been increasingly adopted in many applications \cite{Samuel,Karl2018,Maryam}.

For the above mentioned multi-target filtering solutions, most of them always assume that all of states of all targets follow the same stochastic dynamic model at each time, such as the nearly constant velocity (CV) model or coordinated turn (CT) model \cite{Lixiaorong}. However, in many real-world scenes, it is not good enough to consider only one model. Such situations happen, for example, that if the model used by the filter does not match the actual system dynamics in the single-model system, the tracking performance will tend to diverge. Hence, the target motion uncertainty should be also considered. Fortunately, the multiple model (MM) has been proposed for this defect and a popular approach in the literature is the jump Markov system (JMS) \cite{BarshalomBook,Blom,Mahler2012} where target state is augmented with an additional motion model parameter, and the augmented state of each target evolves with time (including the prediction and update processes) via a finite state Markov chain (MC) \cite{Mahler2012}.

A closed-form PHD filter under the linear JMS is proposed in \cite{Pasha2006}, and the further unscented transform technique and linear fractional transformation are combined with the GM-PHD filter for the non-linear JMM \cite{Pasha2009}. Meanwhile, the MM-PHD filter implemented by SMC approach is proposed in \cite{Puni2008}, which can solve the non-Gaussian and non-linear model. Considering the inaccuracy in cardinality of the MM-PHD filter, MM-CPHD filter is proposed in \cite{Georgescu2011,Jie2014}. Although the MM-CPHD filter can improve the performance both states and cardinality, the computational burden increases markedly. Subsequently, the MM-MB filter is proposed in \cite{Dunne2013,Jinlong2012} by propagating the approximated posterior density. Furthermore, some similar ideas based on labeled RFS filters are also proposed in \cite{Punchihewa,Meng2016,Wei2017}.

Inspired by the conjugacy property and low detection tolerance of the PMBM filter, here we address the multiple model PMBM (PMBM) filter to track the maneuvering targets. Meanwhile, considering the superiority of the MM-MB filter than MM-PHD and MM-CPHD filters, we compare the proposed MM-PMBM filter with the MM-MB filter in order to analyze its performance and characteristics. The contribution of the paper is mainly as follows:
\begin{enumerate}
  \item The explicit formulas for the proposed MM-PMBM filter, including prediction and update steps, are provided. Meanwhile, based on the Gaussian mixture implementation, the analytic implementation of the MM-PMBM filter based on linear Gaussian JMS is derived. Furthermore, the comparison between the MM-PMBM and MM-MB filters is provided via simulations, including the states errors and cardinality estimation. In addition, some comparisons with variable detection probability and measurement noise are also provided.
\end{enumerate}

The outline of the rest of the paper is as follows. Section II introduces some background knowledge on RFS, and Section Section III introduces the JMS and MM-PMBM filter in theory. Section IV details the implementation of proposed MM-PMBM filter based on Gaussian mixture. Simulation results are provided in Section V, and conclusions are drawn in Section VI.


\section{Background on RFS}
In this section, some nations are given in \emph{Tabel I}. Moreover, four RFS types, the Poisson RFS, Bernoulli RFS, and Multi-Bernoulli mixture RFS as well as the PMBM filter are successively introduced. More details on these mentioned RFSs can be found in \cite{Mahlerbook,Angel2018}.
\begin{table}[htp!]
  \begin{center}
     \caption{NOMENCLATURE} \label{table0}
\begin{tabularx}{\linewidth}{l @{\hspace{1.5em}=\hspace{1em}}X}
  \toprule
  \toprule
  \multicolumn{1}{l}{\textbf{Abbreviations:}} &
  \multicolumn{1}{@{}X}{\textbf{Complete expression}}\\%
  \midrule
  MTT       & Multi-target tracking\\
  FISST     & Finite set statistics \\
  RFS       & Random finite set                 \\
  JPDA      & Joint probabilistic data association \\
  MHT       & Multiple hypothesis tracking \\
  PPP       & Poisson point process          \\
  PHD       & Probability hypothesis density \\
  CPHD      & Cardinality PHD \\
  MB        & Multi-Bernoulli  \\
  MBM       & Multi-Bernoulli mixture \\
  PMBM      & Poisson multi-Bernoulli mixture \\
  LMB       & Labeled multi-Bernoulli \\
  MM        & Multiple model \\
  IMM       & Interacting multiple model \\
  JMS       & Jump Markov system \\
  CT        & (Nearly) constant turn \\
  CV        & (Nearly) constant velocity \\
  TPM       & Transition probability matrix \\
  GM        & Gaussian mixture \\
  SMC       & Sequential monte carlo \\
  OSPA      & Optimal SubPattern Assignment \\[0.3ex]%
  \midrule
  \multicolumn{1}{l}{\textbf{Mathematical symbols:}} &
  \multicolumn{1}{@{}X}{\textbf{Explanation}}\\%
  \midrule
  $x$                     & target state \\
  $\tilde x$              & augmented target state \\
  $X$                     & RFS of target   \\
  $z$                     & measurement state \\
  $Z$                     & measurement set \\
  $\mathbb{X}$            & state space \\
  ${\mathbb I}$           & index set \\
  $\left| \cdot \right|$  & set cardinality \\
  $p_S $                  & survival probability \\
  $p_D$                   & detection probability \\
  $\lambda$               & clutter rate \\
  $\xi$                   & motion model \\
  $\mu$                   & first-order moment \\
  ${\cal N}(\cdot)$       & Gaussian function \\
  $\omega$                & hypothesis weight \\
  $r$                     & existing probability \\
  $f(\cdot)$              & probability density function \\[0.3ex]%
  \bottomrule
  \bottomrule
\end{tabularx}
\end{center}
\end{table}

\subsection{Poisson RFS}

Poisson point process (PPP) is parameterized by its intensity function or first-order moment $\mu(x)=\lambda f(x)$, where $\lambda$ is the Poisson rate and $f(x)$ is a probability density function (pdf) of single target, meanwhile, the cardinality of PPP follows a Poisson distribution and its element obeys independently and identically distributed (i.i.d.). The corresponding multi-target density of a Poisson RFS is
\begin{eqnarray}
\label{PPP1}
f(X) &=& {e^\lambda }\prod\limits_{i = 1}^n {\lambda f({x_i})}\\
\label{PPP2}
&=& {e^{ - \int {\mu (x)dx} }}{\left[ {\mu ( \cdot )} \right]^X}.
\end{eqnarray}
\subsection{Bernoulli RFS}
A Bernoulli RFS $X$ can be expressed as follows.
\begin{equation}
f(X) = \left\{ \begin{array}{l}
1 - r,{\kern 1pt} {\kern 1pt} {\kern 1pt} {\kern 1pt} X = \emptyset \\
rf(x),{\kern 1pt} X = \{ x\} \\
0,{\kern 1pt} {\kern 1pt} {\kern 1pt} {\kern 1pt} {\kern 1pt} {\kern 1pt} {\kern 1pt} {\kern 1pt} {\kern 1pt} {\kern 1pt} {\kern 1pt} {\kern 1pt} {\kern 1pt} {\kern 1pt} {\kern 1pt} \left| X \right| \ge 2
\end{array} \right.
\end{equation}
where $r \in [0,1]$ is the existence probability and $f(x)$ denotes the pdf if the target exists.
\subsection{MBM RFS}
An MB RFS is the disjoint union of independent Bernoulli RFS,
\begin{equation}
X = \bigcup\nolimits_{i = 1}^n {{X_{i}}}
\end{equation}
where $n$ denotes the number of Bernoulli component. Moreover, the MB distribution can be completely characterized by a set of parameters $\left\{ {{r_i},{f_i}(x)} \right\}_{i = 1}^n$, which can be expressed as follows.
\begin{equation}
\label{BM}
{f^{{\mathop{\rm mb}\nolimits}}}(X) \propto {\sum\limits_{{X_1} \cup  \cdots  \cup {X_n} = X} {\prod\limits_{i = 1}^n {{\omega_{i}}{f_{i}}({X_i})} } }
\end{equation}

where $\omega_i$, $r_i$ and $f_i(x)$ denote the weight, existence probabilty and pdf of $i$-th target.

The MBM RFS is the normalized and weighted sum of multi-target densities of MBs, which is parameterized by ${\left\{ {{w_{j,i}}, {{\left\{ {{r_{j,i}},{f_{j,i}}(x)} \right\}}_{i \in {{\mathbb I}^j}}}} \right\}_{j \in {\mathbb I}}}$, where $\mathbb I$ is the index set of the MBs in the MBM. The multi-target distribution is
\begin{equation}
\label{MBM}
{f^{{\mathop{\rm mbm}\nolimits}}}(X) \propto \sum\limits_{j \in {\mathbb I}} {\sum\limits_{{X_1} \cup  \cdots  \cup {X_n} = X} {\prod\limits_{i = 1}^n {{\omega_{j,i}}{f_{j,i}}({X_i})} } }
\end{equation}
where $\propto$ stands for proportionality, and $j$ is an index over all global hypotheses (components of the mixtures). Compared (\ref{BM}) with (\ref{MBM}), it can be seen that MB RFS is a special case of an MBM when $\left| \mathbb I \right| = 1$.

\subsection{PMBM Filter}

For the PMBM filter, conditioned on the measurement set $Z^{1:k}$, the multi-target state RFS $X_k$ at time $k$ is modeled as the union of independent RFS $X_k^u$ (undetected targets\footnote{Undetected target: exists at the current time but have never been detected.}) and $X_k^d$ (potentially detected targets\footnote{Potentially detected target: a new measurement may be a new target for the first detection and can also correspond to another previously detected target or clutter. Considering that it may exist or not, we refer to it as potentially detected target.}), respectively. Hence, the posterior pdf of the PMBM filter can be denoted by the FISST convolution as
\begin{equation}
f\left( X_k \right) = \sum\limits_{Y_k\bigcup W_k  = X_k} {{f^{{\mathop{\rm p}\nolimits}}}(Y_k){f^{{{\mathop{\rm mbm}\nolimits}}}}(W_k)}
\end{equation}
where $f^{{\mathop{\rm p}\nolimits}}(\cdot)$ is a Poisson density which is shown in (\ref{PPP2}) and $f^{{{\mathop{\rm mbm}\nolimits}}}(\cdot)$ is an MBM which is shown in (\ref{MBM}). Next, the recursive process will be given.
\subsubsection{Prediction Steps}
For the prediction process of the PMBM filter, two parts including the Poisson density $f^{\mathop{\rm p}\nolimits}(\cdot)$ and the multi-Bernoulli density $f^{{\mathop{\rm mbm}\nolimits}}(\cdot)$ can be predicted separately.

Suppose the Poisson intensity at time $k-1$ is $\mu_{k-1}(x_{k-1})$, then the prediction of Poisson intensity at time $k$ is
\begin{equation}\label{Pre_poisson}
\begin{split}
&\mu_{k|k-1} (x_k) \\
&= {\lambda ^b}(x_k)+ \int {f(x_k|x_{k-1}){p_S}(x_{k-1}){\mu_{k-1}}(x_{k-1})dx_{k-1}}
\end{split}
\end{equation}
where $\mu_{k|k-1}(x_k)$ is the predicted intensity of Poisson components at time $k$, $\lambda^b(x_k)$ and $\mu_{k-1}(x_{k-1})$ are the intensities of birth model and Poisson components from time $k-1$, respectively. $f(x_k|x_{k-1})$ denotes the state transition function, and $p_S(\cdot)$ survival probability.

For the MB components, given the $i$-th target in $j$-th global hypothesis at time $k-1$, $\left\{ {\omega_{k - 1}^{j,i},r_{k - 1}^{j,i},p_{k - 1}^{j,i}({x_{k - 1}})} \right\}$, then the prediction process is as follows,
\begin{eqnarray}
\label{Pre_bernoulli_w}
\omega_{k|k-1}^{j,i} &=& \omega_{k-1}^{j,i} \\
\label{Pre_bernoulli_r}
{r_{k|k-1}^{j,i}} &=& r_{k-1}^{j,i}\int {p_{k-1}^{j,i}(x_{k-1}){p_S}(x_{k-1})dx_{k-1}} \\
\label{Pre_bernoulli_p}
{p_{k|k-1}^{j,i}}(x_{k}) &\propto& \int {f(x_k|x_{k-1}){p_S}(x_{k-1})p_{k-1}^{j,i}(x_{k-1})dx_{k-1}}
\end{eqnarray}
where $\omega_{k|k-1}^{j,i},r_{k|k-1}^{j,i},p_{k|k-1}^{j,i}(x_k)$ denote the predicted hypothesis weight, existence probability, and pdf of the $i$-th Bernoulli component in the $j$-th global hypothesis, respectively.
\subsubsection{Update Steps}
The update process consists of three parts: a) update for undetected targets; b) update for potential targets detected for the first time; c) update for previously potentially detected targets including misdetection and measurement update (marked as c.1 and c.2, respectively).

{a) Update for undetected targets:}\\
\begin{equation}\label{undetected_Poisson}
{q_k^p}(U) \propto {\left[ {(1 - {p_D(x_k)})\mu_{k|k-1} (x_k)} \right]^U}
\end{equation}
where $U$ represents the set of undetected targets and $p_D(\cdot)$ is the detection probability.

{b) Update for potential targets detected for the first time:}\\
\begin{eqnarray}
\label{potential_MB_r}
{r_k^p}(z_k) &=& {{e_k(z_k)} \mathord{\left/  \right.} {{\rho_k ^p}(z_k)}}\\
\label{potential_MB_p}
{p_k^p}(x_k|z_k) &=& {p_D(x_k)}l(z_k|x_k)\mu_{k|k-1} (x_k)/e_k(z_k)
\end{eqnarray}
where
\begin{eqnarray}
\nonumber
{\rho_k ^p}(z_k) &=& e_k(z_k) + c(z_k)\\
\nonumber
e_k(z_k) &=& \int {l(z_k|x_k){p_D(x_k)}\mu_{k|k-1} (x_k)dx_k}.
\end{eqnarray}

{c.1) Misdetection for previously potentially detected targets\footnote{For In order to distinguish the misdetection and measurement update, here $\{w,r,p\}$ are attached a symbol $\emptyset$.}:}\\

\begin{eqnarray}
\label{miss1}
{\omega_k^{j,i}}(\emptyset) &=& {\omega_{k|k-1}^{j,i}}(1 - {p_D(x_k)}{r_{k|k-1}^{j,i}})\\
\label{miss2}
{{r}_k^{j,i}}(\emptyset) &=& {r_{k|k-1}^{j,i}}(1 - {p_D(x_k)})/(1 - p_D(x_k){r_{k|k-1}^{j,i}})\\
\label{miss3}
p_k^{j,i}(\emptyset) &=& p_{k|k-1}^{j,i}(x_k).
\end{eqnarray}

{c.2) Update for previously potentially detected targets using measurements:}\\
\begin{equation}
\begin{split}
{{\omega}_k^{j,i}}(z_k) =& {\omega_{k|k-1}^{j,i}}{r_{k|k-1}^{j,i}} \int {{p_D(x_k)}l(z_k|x_k){p_{k|k-1}^{j,i}}(x_k)dx_k} \\
r_k^{j,i}(z_k) =& 1\\
{{p}_k^{j,i}}(x_k,z_k) \propto& {p_D(x_k)}l(z_k|x_k){p_{k|k-1}^{j,i}}(x_k).
\end{split}
\end{equation}

Note that the update steps are also thought separately, update a) comes from Poisson components, and b) and c) are both for the MB RFS. The difference between b) and c) is that the Bernoulli in b) are generated by the predicted Poisson components.

After all single-target hypotheses are done, the global hypotheses should be generated. In theory, a previous global hypothesis must go through all possible data association hypotheses to obtain the updated global hypotheses. Considering the bottleneck of computation, an effective and fast method named \emph{Murty's algorithm} \cite{murty} is adopted, where a cost matrix is constructed using the calculated weight in (14), (16) and (18), and the detailed implementation steps can be referred in \cite{Angel2018}.

\section{MM-PMBM Filter}
In this section, the MM-PMBM filter is introduced. The jump Markov system is given firstly, and then the detailed process of the proposed MM-PMBM filter is provided.
\subsection{Jump Markov System}
Under a Jump Markov modeled system, the motion models of each target switch according to a finite-state Markov chain. In a linear JMS, the probability of transition between two models $f(\xi=n|\xi'=m)$ are constant, which forms a transition matrix $\Pi$. Meanwhile, the random state variable is augmented with motion model. That is $\tilde x = (x,\xi)$. The corresponding pdf of the extended random state can be represented as
\begin{equation}
p(\tilde x) = \int_\Pi  {p(x,\xi)d\xi}  = \sum\limits_n {f(\xi = n)p(x|\xi = n)}.
\end{equation}


The transition density and measurement likelihood for the augmented single-target state can be expressed as
\begin{eqnarray}
\label{transition}
f(x_k,\xi|x_{k-1},\xi') &=& f(x_k|x_{k-1})f_{k|k-1}(\xi|\xi')\\
\label{likelihood}
l(z_k|x_k,\xi) &=& l(z_k|x_k).
\end{eqnarray}
As shown in (\ref{likelihood}), it is typically independent of motion model for the measurement likelihood function.


\subsection{MM-PMBM Filter}
Detailed filtering iterative process is given as follows.
\subsubsection{Prediction Steps}
The Poisson components and multi-Bernoulli components can be predicted separately just like the standard PMBM filter.

{a) Poisson components:} Combine (\ref{Pre_poisson}) and (\ref{transition}), we can obtain the predicted Poisson components.
\begin{equation}
\begin{split}
&\mu_{k|k-1} (x_k,\xi) = {\lambda ^b}(x_k,\xi) \\
&+ \sum\limits_{\xi'} {\int {f(x_k,\xi|x_{k-1},\xi'){p_S}(x_{k-1},\xi'){\mu_{k-1}}(x_{k-1},\xi')dx_{k-1}} }.
\end{split}
\end{equation}

{b) MB components:} The predicted Bernoulli components at time $k$ have the same hypothesis weights as at time $k-1$, $\omega_{k|k-1}^{j,i}=\omega_{k-1}^{j,i}$, which is agreement with (\ref{Pre_bernoulli_w}). The existence probability and pdf in (\ref{Pre_bernoulli_r}) and (\ref{Pre_bernoulli_p}) are augmented with model as follows.
\begin{eqnarray}
{r_{k|k-1}^{j,i}} = r_{k-1}^{j,i}\sum\limits_{\xi'} {\int {p_{k-1}^{j,i}(x_{k-1},\xi'){p_S}(x_{k-1},\xi')dx_{k-1}} }
\end{eqnarray}
\begin{equation}
\begin{split}
&{p_{k|k-1}^{j,i}}(x_k,\xi)\\
 &\propto \sum\limits_{\xi'} {\int {f(x_k,\xi|x_{k-1},\xi'){p_S}(x_{k-1},\xi')p_{k-1}^{j,i}(x_{k-1},\xi')dx_{k-1}} }.
\end{split}
\end{equation}

\subsubsection{Update Steps}
The update process consists of the same three parts as the update process of PMBM filter mentioned in Section II.D.

{a) Update for undetected targets:} The equation (\ref{undetected_Poisson}) is evolved into the following form.
\begin{equation}
{q_k^p}(U,\xi) \propto {\left[ {(1 - {p_D(x_{k},\xi)})\mu_{k|k-1} (x_k,\xi)} \right]^U}.
\end{equation}

{b) Update for potential targets detected for the first time:} The correspodning equations (\ref{potential_MB_r}) and (\ref{potential_MB_p}) are newly denoted as
\begin{eqnarray}
{r_k^p}(z_k) &=& {{e_k(z_k)} \mathord{\left/ \right.} {{\rho_k ^p}(z_k)}}\\
{p_k^p}(x_k,\xi|z_k) &=& \frac{{p_D}(x_{k},\xi)l(z_k|x_k)\mu_{k|k-1} (x_k,\xi)}{e_k(z_k)}
\end{eqnarray}
where
\begin{eqnarray}
\label{weight_potential}
{\rho_k ^p}(z_k)&=& e_k(z_k) + c(z_k)\\
e_k(z_k)&=& \sum\limits_{\xi'} {\int {l(z_k|x_k){p_D(x_k,\xi')}\mu_{k|k-1}(x_k,\xi')dx_k}}.
\end{eqnarray}

{c.1) Misdetection for previously potentially detected targets:}\\

\begin{eqnarray}
\label{GMmiss1}
{{\omega}_k^{j,i}}(\emptyset) &=& {\omega_{k|k-1}^{j,i}}{\rho_k^{j,i}}(\emptyset)\\
\label{GMmiss2}
p_k^{j,i}(\emptyset ,\xi ) &=& {{(1 - {p_D}(\xi ))p_{k|k - 1}^{j,i}({x_k},\xi )} \mathord{\left/
 {\vphantom {{(1 - {p_D}(\xi ))p_{k|k - 1}^{j,i}({x_k},\xi )} {\rho _k^{j,i}(\emptyset )}}} \right.
 \kern-\nulldelimiterspace} {\rho _k^{j,i}(\emptyset )}}
\end{eqnarray}
\begin{equation}\label{GMmiss3}
\begin{split}
&{{r}_k^{j,i}}(\emptyset) \\
&= \frac{{{r_{k|k-1}^{j,i}}{\sum\nolimits_{\xi'} {\int {(1 - {p_D}(x_k,\xi')){p_{k|k-1}^{j,i}}(x_k,\xi')dx_k} } }}}{{{\rho_k^{j,i}}(\emptyset )}}
\end{split}
\end{equation}
where
\begin{equation}
\begin{split}
\label{weight_empty}
{\rho_k^{j,i}}(\emptyset ) =& 1 - {r_{k|k-1}^{j,i}} \\
&+ {r_{k|k-1}^{j,i}}\sum\limits_{\xi'} {\int {(1 - {p_D}(x_k,\xi')){p_{k|k-1}^{j,i}}(x_k,\xi')dx_k} }.
\end{split}
\end{equation}

{c.2) Update for previously potentially detected targets using measurements $z_k$:}\\

\begin{equation}\label{GMupdate1}
\begin{split}
{{\omega}_k^{j,i}}(z_k)=& {\omega_{k|k-1}^{j,i}}{r_{k|k-1}^{j,i}}\\
&\cdot\sum\limits_{\xi'} {\int {{p_D}(x_k,\xi')l(z_k|x_k){p_{k|k-1}^{j,i}}(x_k,\xi')dx_k} }\\
\end{split}
\end{equation}
\begin{eqnarray}
\label{GMupdate2}
r_k^{j,i}(z_k)&=& 1\\
\label{GMupdate3}
{{p}_k^{j,i}}(x_k,\xi,z_k) &\propto& {p_D}(x_k,\xi)l(z_k|x_k,\xi){p_{k|k-1}^{j,i}}(x_k,\xi).
\end{eqnarray}

After we have calculated all possible new single-target hypotheses, the global hypotheses need to form for satisfying the next recursion. In order to avoid all possible data hypotheses for each previous global hypothesis, we still adopt the Murty's algorithm. First, a cost matrix using the updated weights of the conjugate prior should be constructed. Assume there are $n_o$ old tracks in the global hypothesis $j$ and $m$ measurements ${z_k^1,\cdots,z_k^m}$, which indicates that there are $m$ potential detected targets. The cost matrix at time $k$ can be formed as follows.
\begin{equation}\label{cost_matrix}
C_j = -ln \left[ {\begin{array}{*{20}{c}}
{\eta_{j}^{1,1}}&{\eta_{j}^{1,2}}& \cdots &{\eta_p^{1,1}}& \cdots &0\\
 \vdots & \vdots & \vdots & \vdots & \vdots & \vdots \\
{\eta_{j}^{m,1}}&{\eta_{j}^{m,2}}& \cdots &0& \cdots &{\eta_p^{m,m}}
\end{array}} \right]
\end{equation}
where $\eta_{j}^{m,n}$ denotes the weight after $m$-th measurement updates the $n$-th old track in the $j$-th global hypothesis, which is
\begin{equation}\label{weight_update}
\eta_{j}^{m,n} = {\omega_k^{j,i}}{\rho_k^{j,i}}(z_k^m)/{\rho_k^{j,i}}(\emptyset)
\end{equation}
with ${\rho_k^{j,i}}(z_k^m)$ given by
\begin{equation}
\begin{split}
\nonumber
&{\rho_k^{j,i}}({z_k^m} ) = {r_k^{j,i}}\sum\limits_{\xi'} {\int {{p_D}(x_k,\xi')l(z_k^m|x_k){p_{k|k-1}^{j,i}}(x_k,\xi')dx_k} }
\end{split}
\end{equation}
and ${\rho _k^{j,i}}(\emptyset)$ given by (\ref{weight_empty}). Moreover, $\eta_p^{m,m}$ denotes the weight of $m$-th potential detected target given by (\ref{weight_potential}).

\begin{Rem}
The construction method of cost matrix is the same as the standard PMBM filter, because the hypothesis weight of each single target is not related to the motion model. But the difference is that the computation of parameter in the matrix involves the summation of all motion models.
\end{Rem}


\begin{Rem}
From the point of computational burden, it is evident that the MM-PMBM filter is a larger amount of computation than PMBM filter because the interaction process is needed between different motion models. Moreover, to reduce the computational burden, some measurements can be removed by gating \cite{Kurien} before update process just like the processing in the PMBM filter, which can sharply reduce the dimension of cost matrix.
\end{Rem}

\section{Gaussian mixture Implementation}
In this section, we will denote the implementation of MM-PMBM filter based on GM technology \cite{Vo2}. We also assume the augmented state transition density satisfies
\begin{equation}
\nonumber
f(x_k,\xi|x_{k-1},\xi')=f_{k|k-1}(\xi|\xi'){\cal N}\left(x_k;F_k(\xi)x_{k-1},Q_k(\xi)\right)
\end{equation}
where $F_k(\xi)$ denotes the state transition matrix with model $\xi$ and $Q_k(\xi)$ is the covariance of process noise with model $\xi$. The likelihood function is
\begin{equation}
\nonumber
l(z_k|x_k) = {\cal N}\left(z_k; H_kx_k,R_k\right).
\end{equation}
Moreover, the detection and survival probabilities are usually assumed model dependent,
\begin{equation}
\begin{split}
\nonumber
p_D(x_k,\xi) &= p_D(\xi) \\
p_S(x_k,\xi) &= p_S(\xi).
\end{split}
\end{equation}

To derive the closed-form MM-PMBM recursion, two necessary lemmas are stated as follows:
\begin{Lem}
Given $F$, $m$, and positive definite matrix $Q$ and $P$, then
\begin{equation}
\nonumber
\int {{{\cal N}}\left( {x;F\zeta ,Q} \right){{\cal N}}\left( {\zeta ;m,P} \right)d\zeta  = {{\cal N}}\left( {x;Fm,Q + FP{F^T}} \right)}.
\end{equation}
\end{Lem}
\begin{Lem}
Given $H$, $m$, and positive definite matrix $R$ and $P$, then
\begin{equation}
\nonumber
{{\cal N}}\left( {{z};H{x},{R}} \right){{\cal N}}\left( {x;{m},{P}} \right) = q(z){{\cal N}}\left( {x;{\hat m},{\hat P}} \right)
\end{equation}
where
\begin{equation}
\begin{split}
\nonumber
q(z)&={\cal N}(z;Hm,R+HPH^{\top})\\
{\hat m} &= m+K(z-Hm) \\
{\hat P} &= (I-KH)P \\
K &= PH^{\top}(HPH^{\top}+R)^{-1}.
\end{split}
\end{equation}
\end{Lem}
\subsection{Predict Process}
First, the intensity of birth RFS can be expressed as Gaussian mixtures of the form:
\begin{equation}
{\lambda^b}\left( {x_k,\xi} \right) = {f_k^{b}}(\xi)\sum\limits_{q = 1}^{{J_{\gamma ,k}}(\xi)} {w_{\gamma ,k}^{q}(\xi){{\cal N}}\left( {x_k;x_{\gamma ,k}^{q}(\xi),P_{\gamma ,k}^{q}(\xi)} \right)}
\end{equation}
where ${\cal N}\left(\cdot\right)$ denotes the Gaussian kernel, and $w$ denotes the weight of Gaussian kernel. $f_k^b(\xi)$ is the initial transition probability of model $\xi$.

In terms of the Poisson part, suppose the posterior intensity at time $k-1$ is a Gaussian mixture
\begin{equation}
\begin{split}
&{\mu_{k-1}}(x_{k-1},\xi)\\
&= \sum\limits_{q = 1}^{{J_{k-1}^{\mu}}(\xi)} {{w_{k-1}^{q}}(\xi){{\cal N}}\left( {x_{k-1};x_{k-1}^{q}(\xi),P_{k-1}^{q}(\xi)} \right)}.
\end{split}
\end{equation}
Then, the predicted intensity at time $k$ is also a Gaussian mixture using the Lemma 1,
\begin{equation}\label{pre_Poisson}
\begin{split}
\mu_{k|k-1}(x_k,\xi) =& {\lambda ^b}(x_k,\xi) + \sum\limits_{\xi'} \sum\limits_{q = 1}^{{J_{k-1}^{\mu}}(\xi')} w_{k|k-1}^q(\xi|\xi')\\
&\cdot{{\cal N}}\left( {x_k;x_{k|k-1}^{q}(\xi|\xi'),P_{k|k-1}^{q}(\xi|\xi')} \right)
\end{split}
\end{equation}
where
\begin{equation}
\nonumber
\begin{split}
w_{k|k-1}^q(\xi|\xi') &= {w_{k-1}^{q}}(\xi'){{p_S}(\xi')f_{k|k-1}(\xi|\xi')}\\
m_{k|k-1}^{q}(\xi|\xi') &= F_k(\xi)x_{k-1}^q(\xi') \\
P_{k|k-1}^q(\xi|\xi') &= {Q_k(\xi)} + {F_k(\xi)}P_{k-1}^q(\xi')F_k(\xi)^{\top}.
\end{split}
\end{equation}

In terms of the MBM components, assume that the Bernoulli density at time $k-1$ can be expressed as follows.
\begin{equation}
p_{k - 1}^{j,i}({x_k},\xi ) = \sum\limits_{q = 1}^{J_{k - 1}^{j,i}(\xi )} {w_{k - 1}^{j,i,q}(\xi ){{\cal N}}\left( {{x_{k - 1}};x_{k - 1}^{j,i,q}(\xi ),P_{k - 1}^{j,i,q}(\xi )} \right)}.
\end{equation}
Then the predicted Bernoulli components have the same weights. Moreover, existence probability satisfies
\begin{equation}
r_{k|k - 1}^{j,i} = r_{k - 1}^{j,i}\sum\limits_\xi  {\sum\limits_{\xi '} {\sum\limits_{q = 1}^{J_{k - 1}^{j,i}(\xi ')} {f(\xi |\xi '){p_S}(\xi ')w_{k - 1}^{j,i,q}(\xi ')} } } \\
\end{equation}
and predicted density obtained using Lemma 1 is
\begin{equation}\label{pre_Bernoulli}
\begin{split}
&{p_{k|k-1}^{j,i}}(x_k,\xi) \\
&= \sum\limits_{\xi'}\sum\limits_{q = 1}^{J_{k - 1}^{j,i}(\xi ')}w_{k|k-1}^{j,i,q}(\xi|\xi')
{\cal{N}}\left( {x_k; x_{k|k-1}^{j,i,q}(\xi|\xi'),P_{k|k-1}^{j,i,q}(\xi|\xi')} \right)
\end{split}
\end{equation}
where
\begin{equation}
\nonumber
\begin{split}
w_{k|k-1}^{j,i,q}(\xi |\xi ') &= {p_S(\xi')}f_{k|k-1}(\xi|\xi')w_{k-1}^{j,i,q}(\xi') \\
x_{k|k - 1}^{j,i,q}(\xi |\xi ') &= {F_k}(\xi )x_{k - 1}^{j,i,q}(\xi ')\\
P_{k|k - 1}^{j,i,q}(\xi |\xi ') &= {Q_k(\xi) } + {F_k}(\xi )P_{k - 1}^{j,i,q}(\xi '){F_k}{(\xi )^{\top} }.
\end{split}
\end{equation}
\begin{Rem}
Note that the existence probability and hypothesis weight of Bernoulli components are independent of model $\xi$. Moreover, there are $\sum\nolimits_\xi  {\left( {\left| {{J_{r,k}}(\xi )} \right| + \left| {{J_{k - 1}^{\mu}}(\xi )} \right|} \right)}$ GCs for Poisson components and $\left| {{\mathbb I}} \right|\sum\nolimits_j {\sum\nolimits_\xi  {\left| {{{{\mathbb I}}^j}} \right|\left| {J_{k - 1}^{j,i}(\xi )} \right|} } $ GCs for multi-Bernoulli mixture components to be stored at the prediction process of time $k$.
\end{Rem}
\subsection{Update Process}
Rewrite the predicted intensity of the Poisson part as
\begin{equation}
\begin{split}
&\mu_{k|k-1}(x_k,\xi) \\
&= \sum\limits_{q = 1}^{{J_{k|k-1}^\mu }(\xi)} {w_{k|k-1}^{\mu ,q}}(\xi){{\cal N}}\left( {x_k;x_{k|k-1}^{\mu ,q}(\xi),P_{k|k-1}^{\mu ,q}(\xi)} \right).
\end{split}
\end{equation}
Then the updated posterior intensity ${\mu}_k(x_k,\xi)$ is also a Gaussian mixture given by
\begin{equation}
{\mu}_k(x_k,\xi) = (1-p_D)\mu_{k|k-1}(x_k,\xi).
\end{equation}

For the update process of the Bernoulli components, there are two types, update for potential targets detected for the first time and update for previously potentially detected targets. First, we rewrite (\ref{pre_Bernoulli}) as
\begin{equation}
\begin{aligned}
&p_{k|k - 1}^{j,i}({x_k},\xi ) \\
&=\sum\limits_{q=1}^{J_{k|k-1}^{j,i}(\xi)} {w_{k|k-1}^{j,i,q}(\xi)}{{\cal N}}\left( {{x_{k}};x_{k|k - 1}^{j,i,q}(\xi ),P_{k|k - 1}^{j,i,q}(\xi)} \right).
\end{aligned}
\end{equation}

\subsubsection{Potential detected targets for the first time}
The newborn targets come from the measurements, and the existence probability is
\begin{equation}
{r_k^p}(z_k) = e_k(z_k)/{\rho_k^p}(z_k)
\end{equation}
and density is obtained using Lemma 2,
\begin{equation}\label{up_Poisson}
\begin{split}
p_k^p({x_k},\xi ,{z_k}) = \sum\limits_{q = 1}^{J_{k|k - 1}^\mu (\xi )} {w_k^{p,q}(\xi ,{z_k}){{\cal N}}\left( {{x_k};x_k^{p,q}(\xi ,{z_k}),P_k^{p,q}(\xi )} \right)}
\end{split}
\end{equation}
where
\begin{equation}\label{potential_targets_weight}
\begin{split}
e_k(z_k) =& \sum\limits_{\xi'} \sum\limits_{q = 1}^{{J_{k|k-1}^\mu }(\xi')} {p_D(\xi')}{w_{k|k-1}^{\mu ,q}}(\xi')\\
&\cdot {{\cal N}}\left( {z_k;H_k{x_{k|k-1}^{\mu ,q}}(\xi'),{S_k^{\mu ,q}}(\xi')} \right) \\
{\rho_k ^p}(z_k) =& e_k(z_k) + c(z_k)\\
{x_k^{p ,q}}(z_k,\xi) =& {x_{k|k-1}^{\mu,q}}(\xi)\\
&+ {\psi_k^{\mu ,q}}(\xi){\left[ {{S_k^{\mu ,q}}(\xi)} \right]^{ - 1}}\left( {z_k - H_k{x_{k|k-1}^{\mu,q}}(\xi)} \right)\\
x_k^{p,q}(\xi) =& H_kx_{k|k-1}^{p,q}(\xi) \\
P_k^{p ,q}(\xi) =& {P_{k|k-1}^{\mu ,q}}(\xi) - {\psi_k^{\mu ,q}}(\xi){\left[ {{S_k^{\mu ,q}}(\xi)} \right]^{ - 1}}{\left[ {{\psi_k^{\mu ,q}}(\xi)} \right]^{\top}}\\
{\psi_k^{\mu ,q}}(\xi) =& {P_{k|k-1}^{\mu ,q}}(\xi){H_k^{\top}}\\
{S_k^{\mu ,q}}(\xi) =& H_k{P_{k|k-1}^{\mu ,q}}(\xi){H_k^{\top}} + R_k\\
\end{split}
\end{equation}
\begin{equation}
\begin{split}
\nonumber
w_k^{p,q}(\xi ,{z_k})=& \frac{{w_{k|k - 1}^{p,q}(\xi ){p_D}(\xi )C_k(\xi,z_k)}}{{\sum\limits_{\xi '} {\sum\limits_{q = 1}^{J_{k|k - 1}^\mu (\xi ')} {w_{k|k - 1}^{p,q}(\xi '){p_D}(\xi ')C_k(\xi,z_k)} } }}\\
C_k(\xi,z_k) =& {{\cal N}}\left( {{z_k};x_{k}^{p,q}(\xi ),P_k^{p,q}(\xi )} \right).
\end{split}
\end{equation}

\subsubsection{Previously potentially detected targets}

Given an $i$-th predicted singe-target state in the $j$-th global hypothesis at time $k$, marked as $\left\{ {{\omega_{k|k-1}^{j,i}},{r_{k|k-1}^{j,i}},{p_{k|k-1}^{j,i}}(x_k,\xi)} \right\}$, where ${p_{k|k-1}^{j,i}}(x_k,\xi)$ is given in (44). 
This single target state will go through two states: misdetection and update by measurements. \\
For the misdetection, the GM expressions of (\ref{GMmiss1}), (\ref{GMmiss2}) and (\ref{GMmiss3}) are given by
\begin{equation}\label{Denominator}
\begin{split}
&\omega_k^{j,i}(\emptyset) = {\omega_{k|k-1}^{j,i}}\left({1 - r_{k|k - 1}^{j,i}}\right)\\
&+{{\omega_{k|k-1}^{j,i}} r_{k|k - 1}^{j,i}\sum\limits_\xi  {\sum\limits_{q = 1}^{J_{k|k - 1}^{j,i}(\xi )} {(1 - {p_D}(\xi ))w_{k|k - 1}^{j,i}(\xi )} } } \\
\end{split}
\end{equation}
\begin{equation}
\begin{split}
&r_k^{j,i}(\emptyset) \\
&= \frac{{r_{k|k - 1}^{j,i}\sum\limits_\xi  {\sum\limits_{q = 1}^{J_{k|k - 1}^{j,i}(\xi )} {(1 - {p_D}(\xi ))w_{k|k - 1}^{j,i}(\xi )} } }}{{1 - r_{k|k - 1}^{j,i}+ r_{k|k - 1}^{j,i}\sum\limits_\xi  {\sum\limits_{q = 1}^{J_{k|k - 1}^{j,i}(\xi )} {(1 - {p_D}(\xi ))w_{k|k - 1}^{j,i}(\xi )} } }}\\
\end{split}
\end{equation}
\begin{equation}
\begin{split}
&p_k^{j,i}(\emptyset ,\xi ) = \sum\limits_{q = 1}^{J_{k|k - 1}^{j,i}(\xi )} {w_{k}^{j,i,q}(\xi ){{\cal N}}\left( {{x_k};x_{k}^{j,i,q}(\xi ),P_{k}^{j,i,q}(\xi )} \right)}
\end{split}
\end{equation}
where
\begin{equation}
\begin{split}
\nonumber
w_k^{j,i,q}(\xi ) &= \frac{{(1 - {p_D}(\xi ))w_{k|k - 1}^{j,i,q}(\xi )}}{{\sum\nolimits_{\xi '} {\sum\nolimits_{q = 1}^{J_{k|k - 1}^{j,i}(\xi ')} {(1 - {p_D}(\xi '))w_{k|k - 1}^{j,i,q}(\xi ')} } }}\\
x_k^{j,i,q}(\xi ) &= x_{k|k - 1}^{j,i,q}(\xi )\\
P_k^{j,i,q}(\xi ) &= P_{k|k - 1}^{j,i,q}(\xi ).
\end{split}
\end{equation}
For the update by measurement $z_k$, the corresponding GM expressions of (\ref{GMupdate1}), (\ref{GMupdate2}), and (\ref{GMupdate3}) are shown as follows.
\begin{equation}\label{Numerator}
\begin{split}
\omega_k^{j,i}(z_k) =& {\omega_{k|k-1}^{j,i}}{r_{k|k-1}^{j,i}}\\
&\cdot\sum\nolimits_{\xi}\sum\nolimits_{q=1}^{J_{k|k-1}^{j,i}(\xi)}{p_D(\xi)}{w_{k|k-1}^{j,i,q}(\xi)}C_k(\xi,z_k)\\
r_k^{j,i}(z_k) =& 1\\
\end{split}
\end{equation}
\begin{equation}\label{up_Bernoulli}
\begin{split}
&p_k^{j,i}({x_k},\xi,z_k ) \\
&= \sum\limits_{q = 1}^{J_{k|k - 1}^{j,i}(\xi )} {w_k^{j,i,q}(\xi ,{z_k})} {{\cal N}}\left( {{x_k};x_k^{j,i,q}(\xi ,{z_k}),P_k^{j,i,q}(\xi )} \right)
\end{split}
\end{equation}
where
\begin{equation}
\begin{split}
\nonumber
x_k^{j,i,q}(\xi,z_k) =& {x_{k|k-1}^{j,i,q}}(\xi) \\
&+ {\psi_k^{j,i,q}}(\xi){\left[ {{S_k^{j,i,q}}(\xi)} \right]^{ - 1}}\left(z_k - H_k{m_{k|k-1}^{j,i,q}}(\xi)\right)\\
\end{split}
\end{equation}
\begin{equation}
\begin{split}
\nonumber
P_k^{j,i,q}(\xi ) =& P_{k|k - 1}^{j,i,q}(\xi ) - \psi _k^{j,i,q}(\xi )S_k^{j,i,q}(\xi ){\left[ {\psi _k^{j,i,q}(\xi )} \right]^{\top}}\\
{\psi_k^{j,i,q}}(\xi) =& {P_{k|k-1}^{j,i,q}}(\xi){H_k^{\top}}\\
{S_k^{j,i,q}}(\xi) =& H_k{P_{k|k-1}^{j,i,q}}(\xi){H_k^{\top}} + R_k\\
w_k^{j,i,q}(\xi,{z_k})
=& \frac{{{p_D}(\xi )w_{k|k - 1}^{j,i,q}(\xi )C_k(\xi,z_k)}}{{\sum\nolimits_{\xi'} {\sum\nolimits_{q = 1}^{J_{k|k - 1}^{j,i}(\xi')} {{p_D}(\xi ')w_{k|k - 1}^{j,i,q}(\xi ')C_k(\xi,z_k)} } }}\\
C_k(\xi,z_k)=&{{\cal N}}\left( {{z_k};{H_k}x_{k|k - 1}^{j,i,q}(\xi ),S_k^{j,i,q}(\xi )} \right).
\end{split}
\end{equation}

Two important equations (\ref{weight_potential}) and (\ref{weight_update}) are needed when the cost matrix is constructed. The Gaussian form of $e_k(z_k)$ in (\ref{weight_potential}) is shown in (\ref{potential_targets_weight}), and the numerator and denominator in (\ref{weight_update}) are given in (\ref{Numerator}) and (\ref{Denominator}), respectively. Hereafter, the global hypothesis can be obtained by assignment operation in (\ref{cost_matrix}).

\begin{Rem}
Because only the misdetection of Poisson part is considered, the GCs of Poisson part is $\sum\nolimits_\xi  {\left| {J_{k|k - 1}^\mu (\xi )} \right| \cdot \left| {{Z_k}} \right|} $, and the number of GCs of MBM consists of two parts, $\sum\nolimits_\xi  {\left| {J_{k|k - 1}^\mu (\xi )} \right| \cdot \left| {{Z_k}} \right|} + \left| {{\mathbb I}} \right|\sum\nolimits_j {\sum\nolimits_\xi  {\left| {{{{\mathbb I}}^j}} \right|\left| {J_{k|k - 1}^{j,i}(\xi )} \right|} } \left( {\left| {{Z_k}} \right| + 1} \right)$¡£ Thus, the total GCs of the MM-PMBM filter at time $k$ is $\left( {\sum\nolimits_\xi  {\left| {J_{k|k - 1}^\mu (\xi )} \right|}  + \left| {{\mathbb I}} \right|\sum\nolimits_j {\sum\nolimits_\xi  {\left| {{{{\mathbb I}}^j}} \right|\left| {J_{k|k - 1}^{j,i}(\xi )} \right|} } } \right)\left( {\left| {{Z_k}} \right| + 1} \right)$.
\end{Rem}

\begin{Rem}
Though the probabilities of survival and detection should be treated as a random varibale, they are always assumed as the constant in some practical cases. For convinence, here we also assume all the models are matched with the same probabilities of survival and detection in the simulation, that is $p_D(\xi)=p_D$ and $p_S(\xi)=p_S$.
\end{Rem}

\section{Simulation results}

In this section, we implement the performance of the proposed algorithm using GM technology, and examine it in terms of the Optimal SubPattern Assignment (OSPA) errors \cite{Schumacher}. In addition, here we compare the peformance between the MM-PMBM and MM-MB filters for better analysis.

Consider a two-dimensional scenario space of $5000m$ in each dimension, where three targets manoeuver in the surveillance area. Targets 1, 2 and 3 enter the scene at times $k=1,1,10s$ and targets 1 and 2 exit the scene at times $k=40,50s$. The target states consists of 2-dimension position and velocity, i.e., $x_k = \left[ {\begin{array}{*{20}{c}}
{{x_k^1}}&{{{\dot x}_k^1}}&{{x_k^2}}&{{{\dot x}_k^2}}
\end{array}} \right]$. Each target moves independently according to the its motion model $\xi$, i.e.,
\begin{equation}
\nonumber
x_k = {F_k(\xi)}x_{k-1} + u_k
\end{equation}
where $u_k \sim {{\cal N}}\left( {0,\sigma _v^2{{I} _2}} \right)$ with $\sigma_v = 2m$ and $I_n$ denotes the $n \times n$ identity matrice. Meanwhile, measurements, sampled at the observated time $T=60s$, is a noisy verison of the planar position, where observation matrix and measurement noise covariance matrix given by
 \begin{equation}\label{measurement_equation}
 {H_k} = \left[ {\begin{array}{*{20}{c}}
1&0&0&0\\
0&0&1&0
\end{array}} \right],{R_k} = \sigma _\varepsilon ^2{I_2}
\end{equation}
where $\sigma_\varepsilon =10m$ denotes the standard deviation of measurement noise. The probability of survival for each target is $P_S = 0.99$. Clutter is modeled as a Poisson RFS with clutter rate $\lambda_c=10$.

Moreover, each target may randomly change the dynamics of its maneuvers among three possible motion models. Model 1, $Mo_1$, is a CV model with standard deviation of the process noise $\delta_{Mo_1}=5m/s^2$. Model 2, $Mo_2$, is a CT model with standard deviation of the process noise $\delta_{Mo_2}=5m/s^2$ and a counterclockwise turn rate of $10^o/s$. Model 3, $Mo_3$, is a CT model with standard deviation of the process noise $\delta_{Mo_3}=5m/s^2$ and a clockwise turn rate of $10^o/s$.

The linear state transition matrices for the CV and CT models are as follows.
\begin{equation}
\nonumber
{F_{NCV}} = \left[ {\begin{array}{*{20}{c}}
1&T&0&0\\
0&1&0&0\\
0&0&1&T\\
0&0&0&1
\end{array}} \right]
\end{equation}

\begin{equation}
\nonumber
{F_{CT}} = \left[ {\begin{array}{*{20}{c}}
1&{(\sin \theta T)/\theta }&0&{ - (1 - \cos \theta T)/\theta }\\
0&{\cos \theta T}&0&{ - \sin \theta T}\\
0&{(1 - \cos \theta T)/\theta }&1&{(\sin \theta T)/\theta }\\
0&{\sin \theta T}&0&{\cos \theta T}
\end{array}} \right]
\end{equation}

\begin{equation}
\nonumber
{Q_{CV}}{\rm{ = }}\sigma _{CV}^2\left[ {\begin{array}{*{20}{c}}
{{{{T^4}} \mathord{\left/
 {\vphantom {{{T^4}} 4}} \right.
 \kern-\nulldelimiterspace} 4}}&{{{{T^3}} \mathord{\left/
 {\vphantom {{{T^3}} 2}} \right.
 \kern-\nulldelimiterspace} 2}}&0&0\\
{{{{T^3}} \mathord{\left/
 {\vphantom {{{T^3}} 2}} \right.
 \kern-\nulldelimiterspace} 2}}&{{T^2}}&0&0\\
0&0&{{{{T^4}} \mathord{\left/
 {\vphantom {{{T^4}} 4}} \right.
 \kern-\nulldelimiterspace} 4}}&{{{{T^3}} \mathord{\left/
 {\vphantom {{{T^3}} 2}} \right.
 \kern-\nulldelimiterspace} 2}}\\
0&0&{{{{T^3}} \mathord{\left/
 {\vphantom {{{T^3}} 2}} \right.
 \kern-\nulldelimiterspace} 2}}&{{{{T^3}} \mathord{\left/
 {\vphantom {{{T^3}} 2}} \right.
 \kern-\nulldelimiterspace} 2}}
\end{array}} \right]
\end{equation}

\begin{equation}
\nonumber
{Q_{CT}}{\rm{ = }}\sigma _{CT}^2\left[ {\begin{array}{*{20}{c}}
{{{{T^4}} \mathord{\left/
 {\vphantom {{{T^4}} 4}} \right.
 \kern-\nulldelimiterspace} 4}}&{{{{T^3}} \mathord{\left/
 {\vphantom {{{T^3}} 2}} \right.
 \kern-\nulldelimiterspace} 2}}&0&0\\
{{{{T^3}} \mathord{\left/
 {\vphantom {{{T^3}} 2}} \right.
 \kern-\nulldelimiterspace} 2}}&{{T^2}}&0&0\\
0&0&{{{{T^4}} \mathord{\left/
 {\vphantom {{{T^4}} 4}} \right.
 \kern-\nulldelimiterspace} 4}}&{{{{T^3}} \mathord{\left/
 {\vphantom {{{T^3}} 2}} \right.
 \kern-\nulldelimiterspace} 2}}\\
0&0&{{{{T^3}} \mathord{\left/
 {\vphantom {{{T^3}} 2}} \right.
 \kern-\nulldelimiterspace} 2}}&{{{{T^3}} \mathord{\left/
 {\vphantom {{{T^3}} 2}} \right.
 \kern-\nulldelimiterspace} 2}}
\end{array}} \right]
\end{equation}
where $T=1s$ is the sampling period.

The Poisson birth intensity is a Gaussian mixture $\lambda^b(x) = \sum\nolimits_{i = 1}^3 {{w_b}{{\cal N}}} \left( {x;m_b^{(i)},{P_b}} \right)$, where ${w_b}=0.1$, $m_b^{(1)} = {\left[ {\begin{array}{*{20}{c}}
{0}&0&{1500}&0
\end{array}} \right]}$, $m_b^{(2)} = {\left[ {\begin{array}{*{20}{c}}
{3000}&0&{1000}&0
\end{array}} \right]}$, $m_b^{(3)} = {\left[ {\begin{array}{*{20}{c}}
{2500}&0&{3000}&0
\end{array}} \right]}$, and ${P_b} = diag([500,100,500,100]^{\top})^2$.
The distribution of the models at birth is taken as
\begin{equation}
{{f_k^b}(\xi)} = \left[ {\begin{array}{*{20}{c}}
{0.5}&{0.25}&{0.25}
\end{array}} \right].
\end{equation}

\begin{figure}
\centering
\includegraphics[width=3.0in]{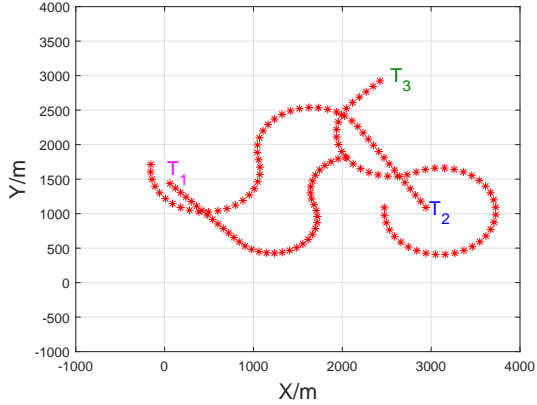}
\caption{The observed region containing three targets, $T_1$, $T_2$ and $T_3$.}
\label{fig:track}
\end{figure}

For all targets, before the time at $k=15s$, they both follow CV motion model if they exist. Once the maneuver happens at time $k=15s$, all targets switch the CV motion model to CT model in counterclockwise direction. At a random time between $k=15s$ and $k=30s$, the motion model is transited to CT model in clockwise direction. The switching between motion models is given by the following Markovian model transition probability matrix (TPM):
\begin{equation}\label{TPM2}
{{f_{k|k - 1}}(\xi|\xi')} = \left[ {\begin{array}{*{20}{c}}
{0.8}&{0.1}&{0.1}\\
{0.1}&{0.8}&{0.1}\\
{0.1}&{0.1}&{0.8}
\end{array}} \right].
\end{equation}
Moreover, the observation region and trajectories are presented in Fig. \ref{fig:track}.

Next, two cases with different detection probability, high $p_D$ and low $p_D$, will be given, where OSPA errors and cardinality are compared. Moreover, the comparison of different $p_D$ and measurement noise are given, respectively. The estimation are averaged over 100 independent Monte Carlo runs.
\subsection{Case 1 with high $p_D$}

\begin{figure}[htbp]
\centering
\includegraphics[width=3.8in]{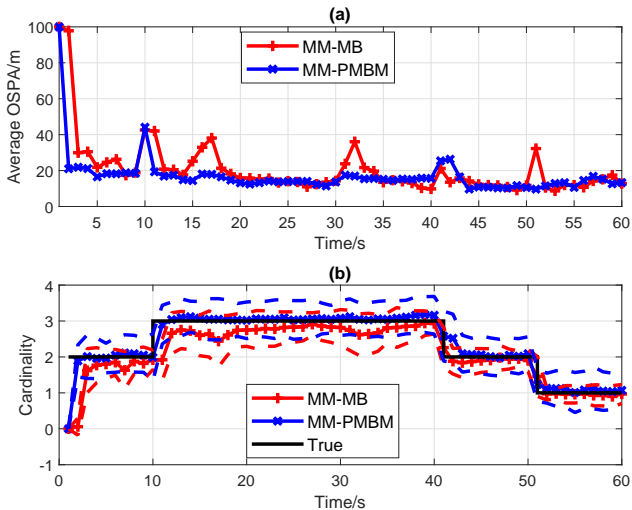}
\caption{The comparison between the MM-BM and MM-PMBM filters with $p_D=0.95$: (a) OSPA errors and (b) cardinality. }
\label{fig:ospa95}
\end{figure}

In this scenario, the detection probability is set $p_D=0.95$, and the simulation results are shown in Fig. \ref{fig:ospa95}. The results demonstrate that in this case, two filters have the similar OSPA errors, but it worth noting that the MM-PMBM filter is not so sensitive compared with the MM-MB filter when the target manoeuvers at $k=15s$ and $t=30s$. However, the MM-PMBM filter is more sensitive if some targets appear ($t=10s$) or disappear ($t=40,50s$). A possible explanation for this phenomenon is that the Poisson part in the PMBM filter retains the components of misdetection and also the components of the target that just disappears, which results in the timely detection of the newborn targets and delayed resposen to death targets. Moreover, the cardinality of PMBM filter is better than the MM-MB filter.

\subsection{Case 2 with low $p_D$}
\begin{figure}[htbp]
\centering
\includegraphics[width=3.8in]{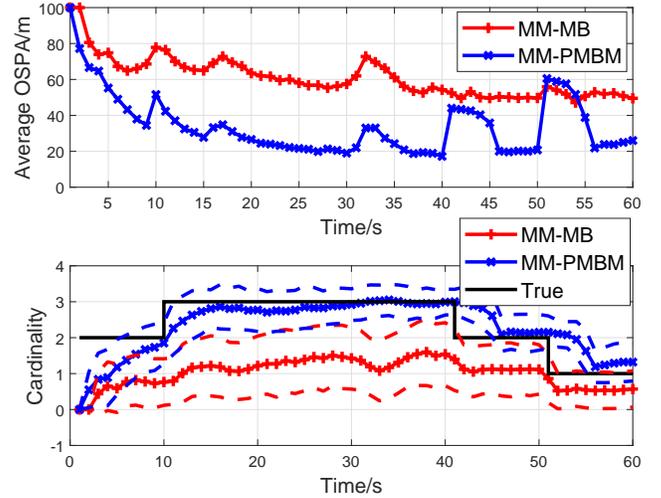}
\caption{The comparison between the MM-BM and MM-PMBM filters with $p_D=0.60$: (a) OSPA errors and (b) cardinality. }
\label{fig:ospa75}
\end{figure}

Diffferent from the case 1, here the low detection probability $p_D=0.60$ is considered. The corresponding results are shown in Fig. (\ref{fig:ospa75}). Compared with the scene with high $p_D$, here the differences of OPSA errors and cardinality between MM-MB and MM-PMBM filters are more obvious, and the MM-PMBM has better performance. It also needs to note that the OSPA errors of the MM-PMBM filter is even worse than MM-MB filter when target disappears. This is because the lower $p_D$ makes the weights of updated Poisson components larger, so that the state of the target that has just disappeared cannot be removed in time.

\subsection{Comparison of different $p_D$}
Changing detection probability $p_D$, we compare the OSPA errors, which can be seen at \emph{Table II}. The resluts show that the proposed MM-PMBM filter has the greater performance advantage at low detection scene.
\begin{table*}[!htbp]
  \begin{center}
     \caption{The comparison about OSPA errors with different detection probabilities $p_D$ ($\sigma_\varepsilon=10m$).} \label{table5}
    \begin{tabular}{ccccccccc}
      \hline
      {\bf\small $p_D$} &{0.60} &{0.65} &{0.70} &{0.75} & {0.80}  & {0.85} & {0.90} & {0.95} \\
      \hline
      \hline
        {MM-MB}   & 68.7989 & 62.5469 & 56.3380 & 47.1008 & 42.4005 & 29.8391 & 25.8295 & 20.9253\\
        \hline
        {MM-PMBM} & 42.2848 & 41.7847 & 40.6930 & 30.5855 & 28.8860 & 27.1736 & 25.2072 & 18.0633\\
        \hline
    \end{tabular}
  \end{center}
\end{table*}
\begin{table*}[!htb]
  \begin{center}
     \caption{The comparison about OSPA errors with different standard deviation of measurement noise in two scenarios, $p_D=0.60/0.95$.} \label{table5}
    \begin{tabular}{cccccc}
      \hline
       {\bf\small $\sigma_\varepsilon$ ($P_D=0.60$)} &{5} & {10}  & {15} & {20} & {25} \\
      \hline
      \hline
        {MM-MB} & 60.7519 & 68.7989 & 75.3410 & 81.0089 & 84.7524\\
        {MM-PMBM} & 37.1890 & 42.2848 & 46.6683 & 49.9221 & 54.9633\\
        \hline
      {\bf\small $\sigma_\varepsilon$($P_D=0.95$)} &{5} & {10}  & {15} & {20} & {25} \\
      \hline
      \hline
        {MM-MB} & 15.2725 & 20.9253 & 28.5589 & 36.2614 & 43.3747\\
        {MM-PMBM} & 16.0537 & 18.0633 & 27.0728 & 32.2202 & 37.1272\\
        \hline
    \end{tabular}
  \end{center}
\end{table*}

\subsection{Comparison of different standard deviation of measurement noise}
Here the standard deviation of measurement noise is changed, and the compared results of two groups are given for high $p_D$ and low $p_D$. From \emph{Tabel III}, the errors of two filters both increase as the standard deviation of measurement noise $\sigma_\varepsilon$ increases. Moreover, the MM-PMBM filter shows its greater advantage in the case of low $p_D$ than the MM-MB filter.

\section{Conclusions}
In this paper, we mainly research the tracking performance of multiple model Poisson multi-Bernoulli mixture (MM-PMBM) filter. In addition to giving the theory, the detailed Gaussian mixture implementation is also provided. We compare the performance, including OSPA errors and cardinality, between the MM-MB filter and the proposed MM-PMBM filter. Furthermore, we also provide the comparison by changing the detection probability and standard deviation of measurement noise, respectively. As a result, the proposed MM-PMBM filter is superior to the MM-MB filter, especially in the scene with low detection probabilty.

As the future work, it would be also interesting to investigate the multi-sensor multiple model filters \cite{Giorgio2015}.

\end{document}